\documentclass[
	pra,
	aps,
	reprint,
	amsmath,amssymb,
	a4paper,
	superscriptaddress,
	10pt
]{revtex4-2}

\usepackage{graphicx}
\usepackage{physics}
\graphicspath{{fig/}}
\usepackage[dvipsnames,table,xcdraw]{xcolor}
\definecolor{rmpblue}{HTML}{2e3092}
\usepackage[
	colorlinks=true,
	citecolor=rmpblue,
	linkcolor=rmpblue,
	filecolor=magenta,
	urlcolor=rmpblue,
]{hyperref}
\usepackage[
	colorlinks=true
]{hyperref}
\usepackage{upgreek}

\usepackage{siunitx}
\newcommand{\affilANU}{Nonlinear Physics Center, Research School of Physics, Australian National University, Canberra ACT 2601, Australia}

\newcommand{\affilEPFL}{Institute of Bioengineering, \'Ecole Polytechnique F\'ed\'erale de Lausanne (EPFL), Lausanne 1015, Switzerland}

\begin{document}

\title{Nonlinear chiral response governed by meta-atom rotation}

\author{Dhruv Hariharan}
\altaffiliation{Contributed equally}
\affiliation{\affilANU}
\author{Pavel Tonkaev}
\email{pavel.tonkaev@anu.edu.au}
\altaffiliation{Contributed equally}
\affiliation{\affilANU}
\author{Felix Ulrich Brikh}
\affiliation{\affilEPFL}
\author{Brijesh Kumar}
\affiliation{\affilANU}
\author{Ivan Toftul}
\affiliation{\affilANU}
\author{Ivan Sinev}
\affiliation{\affilEPFL}
\author{Hatice Altug}
\affiliation{\affilEPFL}
\author{Yuri Kivshar}
\email{yuri.kivshar@anu.edu.au}
\affiliation{\affilANU}

\begin{abstract}
Chiral photonics provides powerful routes for controlling the light handedness, yet nonlinear chiral responses are typically associated with intricate three-dimensional systems. Here, we demonstrate that strong nonlinear chirality can emerge and be precisely tuned in planar metasurfaces. 
We study free-standing membrane metasurfaces composed of periodic lattices of tilted elliptic holes, which preserve out-of-plane mirror symmetry while breaking all in-plane mirror symmetries through the in-plane rotation of the meta-atoms.
We demonstrate that optical resonances play a decisive role in governing the nonlinear chiral response, enabling pronounced circular dichroism in third-harmonic generation even when symmetry is broken only in plane. 
We experimentally reveal strong nonlinear chiral response from the metasurfaces and a striking swapping of nonlinear chiral channels for complementary meta-atom rotation angles. This behaviour arises from the interplay between lattice symmetry and meta-atom orientation, which controls the symmetry of the resonant modes and the resulting nonlinear selection rules. Our results establish meta-atom rotation as a powerful mechanism for engineering nonlinear chiral responses in planar metasurfaces, opening new opportunities for tunable chiral nonlinear metaphotonics devices.

\end{abstract}

\maketitle

\section{Introduction}

Chirality is a fundamental concept in physics defined by non-identity of an object and its mirror image, and it is ubiquitous across science and nature~\cite{prelog_chirality_1976, creutz_aspects_2001,barron_chirality_2008,liu2023detection}. While the qualitative distinction between left-handed and right-handed enantiomers is often straightforward to recognise, establishing a rigorous quantitative measure of chirality remains challenging, as does achieving deterministic control over chiral responses~\cite{mun2020electromagnetic, im2024perspectives,hu2022plasmonic}. In optics, left- and right-circularly polarized electromagnetic waves provide ideal probes of chirality. The interaction of chiral objects with circularly polarized light manifests itself in transmission, reflection, and absorption responses, giving rise to circular dichroism and circular birefringence. These phenomena are widely exploited in many applications ranging from molecular sensing~\cite{hembury_chirality-sensing_2008} and pharmaceutical analysis~\cite{bertucci_circular_2010} to the study of chiral excitations in solid-state~\cite{kuroda_solid-state_2001} and quantum systems~\cite{lodahl2017chiral}.

In recent years, metasurfaces emerged as a powerful platform for tailoring light-matter interaction at the nanoscale \cite{chen_review_2016, krasnok_nonlinear_2018, solntsev_metasurfaces_2021,tonkaev2022multifunctional}. In particular, metasurfaces provide a promising route for controlling chiral optical responses, as their geometry can be precisely engineered to enhance the differential interaction of structures with left- and right-circularly polarized light \cite{shi_planar_2022, shen_chiral_2022, khaliq_recent_2023}. Consequently, chiral optical metasurfaces have rapidly evolved into an active field of research, enabling the control and manipulation of light through engineered optical angular momentum. Over the past decade, extensive efforts have advanced this rapidly developing field, leading to compact chiral metasurface devices and a variety of emerging applications, including chiral sensing \cite{chen_integrated_2020, yoo_metamaterials_2019},  selective chiral transmission~\cite{sinev2025chirality,yang2019spin, kuhner2023unlocking,gryb2023two, duan_spin-preserving_2024}, and chiral lasing \cite{lin_toward_2023, deng_chiral_2025}.

Chiral harmonic generation from nonlinear metasurfaces represents one of the prominent recent developments in chiral metaphotonics \cite{koshelev_nonlinear_2023, lai2025nonlinear}. Nonlinear optical processes provide a powerful route to amplify and probe chiral light-matter interactions, enabling phenomena that are inaccessible in the linear regime. In particular, the presence of resonantly enhanced chiral modes can lead to pronounced nonlinear chiroptical responses, including circular dichroism and helicity-selective harmonic generation~\cite{shi_planar_2022, tonkaev_nonlinear_2024, toftul_chiral_2024,toftul_monoclinic_2025,heimig2026chiral}. Recently, free-standing dielectric membrane metasurfaces have emerged as a particularly attractive platform for exploring such effects, as the absence of a substrate enables strong electromagnetic confinement, improved symmetry control, and enhanced nonlinear interactions. These ultrathin membrane structures have only recently become experimentally accessible and are rapidly finding applications in areas such as optical sensing~\cite{adi2024trapping,rosas_enhanced_2025}, high harmonic generation~\cite{tonkaev_nonlinear_2025}, and compact light-modulation devices~\cite{brikh2025mid}. Remarkably, nonlinear chiral effects can arise even in planar nanostructures that preserve out-of-plane mirror-reflection symmetry. In these systems, the interplay between resonant field confinement, lattice symmetry, and symmetry of individual meta-atoms can give rise to effectively nonreciprocal chiral modes, allowing strong nonlinear chiral responses without requiring intrinsically three-dimensional chiral geometries~\cite{tonkaev_nonlinear_2025}. Despite these advances, the fundamental relationship between nonlinear chiral responses and the combined symmetries of the lattice and the constituent meta-atoms remains poorly understood.

\begin{figure}[h]
    \begin{center}
    \includegraphics[width=0.99\linewidth]{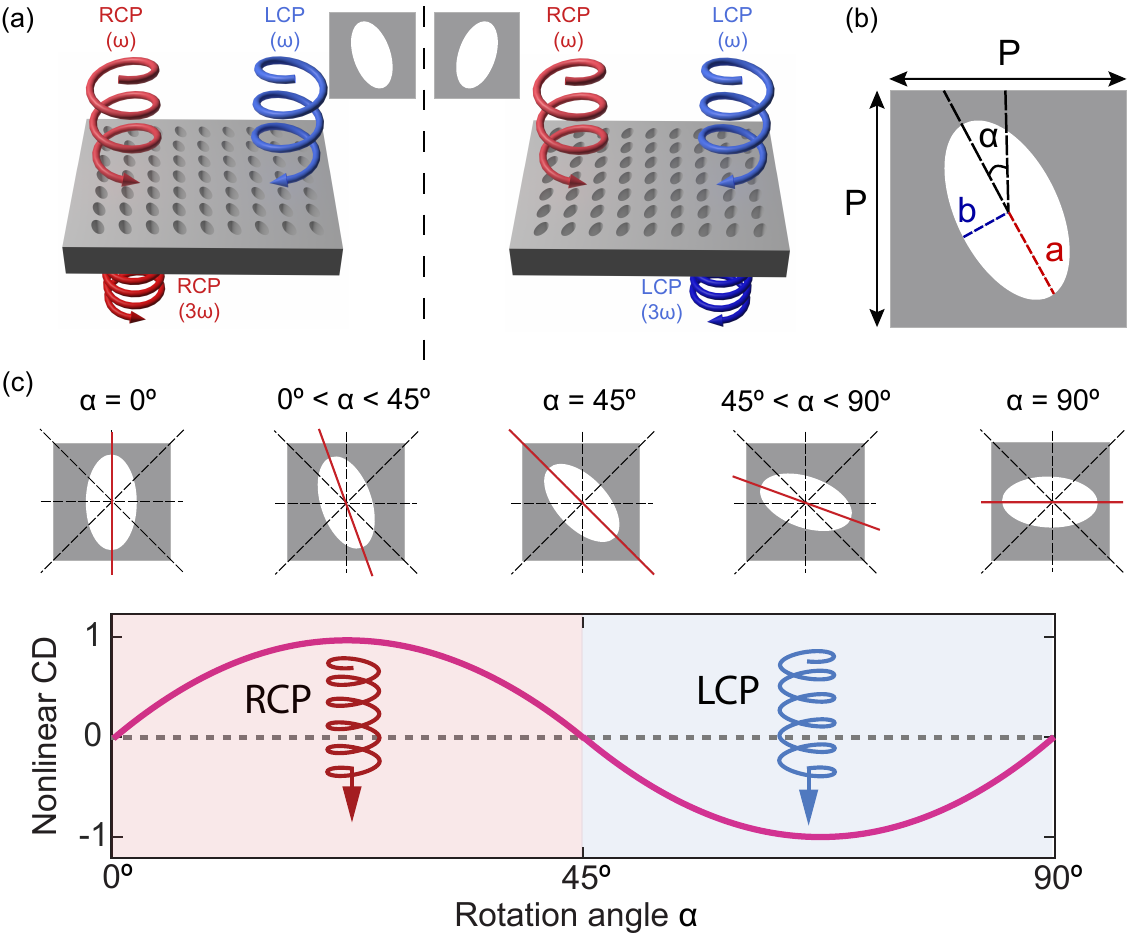}
    \caption{ \textbf{Concept of chiral nonlinear metasurfaces.} (a) Example operation: for a given meta-atom orientation, third-harmonic response is dominated by right-to-right circular polarization conversion, while for the complementary orientation, it is dominated by left-to-left circular polarization conversion. (b) Schematic of an elliptical unit cell with period $P=1500$~nm, the major axis radius $a=610$~nm and the minor axis radius $b=325$~nm. The metasurface thickness is $h=1000$~nm. (c) Nonlinear circular dichroism of the metasurface depends on the symmetry of the unit cell.}
    \label{fig:metasurface_design}
    \end{center}
\end{figure}

In this work, we investigate the nonlinear chiral response of resonant dielectric free-standing membrane metasurfaces composed of periodic lattices of tilted elliptic holes. Such planar structures preserve out-of-plane mirror-reflection symmetry for electromagnetic waves, and are therefore conventionally regarded as achiral. Here, we demonstrate that resonances play a critical role in defining the chiral optical response, even when the structure symmetry is broken only in the plane of the metasurface. We experimentally observe a strong nonlinear chiral response in these resonant dielectric membrane metasurfaces, and reveal a swapping of the nonlinear chiral responses for complementary meta-atom rotation angles [see Figure~\ref{fig:metasurface_design}(a)]. Furthermore, by systematically varying the rotation angle of the elliptical meta-atoms, we observe a transition of the nonlinear circular dichroism from positive to negative values, providing a route to continuously tune the nonlinear circular dichroism to a desired magnitude through meta-atom rotation. This behaviour originates from the interplay between lattice symmetries and meta-atom orientation, highlighting a previously unexplored mechanism for controlling nonlinear chirality in planar metasurfaces.

\section{Results}
\subsection{Metasurface design}

The studied metasurface consists of a 1~\textmu m thick silicon membrane patterned with angled elliptical holes, where a simple design was used to support robust eigenmodes (see Supplementary Section S2). The unit cell is shown in Figure~\ref{fig:metasurface_design}(b), where the ellipse angle $\alpha$ is varied from $0^\circ$ to $90^\circ$ in $1.5^\circ$ increments. The geometric parameters $a = 610$~nm and $b = 325$~nm were optimised using COMSOL Multiphysics for operation with input wavelengths from $3700$~nm to $4000$~nm.

To characterise the linear chirality of the system, we define the co-polarized circular dichroism ($\mathrm{CD}_{\text{co}}$) and total circular dichroism ($\mathrm{CD}_{\text{tot}}$) for their known symmetry selection rules~\cite{koshelev_scattering_2024}:
\begin{equation}\label{eq:CD_lin}
\begin{gathered}
    \mathrm{CD}_{\text{co}} = \frac{T^{( \omega)}_{\text{RR}} - T^{( \omega)}_{\text{LL}}}{T^{( \omega)}_{\text{RR}} + T^{( \omega)}_{\text{LL}}},\\
    \mathrm{CD}_{\text{tot}} = \frac{T^{( \omega)}_{\text{RR}}+T^{( \omega)}_{\text{RL}} - T^{( \omega)}_{\text{LL}}-T^{( \omega)}_{\text{LR}}}{T^{( \omega)}_{\text{RR}}+T^{( \omega)}_{\text{RL}} + T^{( \omega)}_{\text{LL}}+T^{( \omega)}_{\text{LR}}},
\end{gathered}
\end{equation}

where $T^{( \omega)}_{\text{LL}}$ and  $T^{( \omega)}_{\text{RR}}$ denote the co-polarized transmission coefficient for right- and left-circularly polarized (RCP and LCP) incident light, respectively. The terms $T^{( \omega)}_{\text{RL}}$ and  $T^{( \omega)}_{\text{LR}}$ represent cross-polarized transmission coefficients corresponding to conversion from RCP to LCP and from LCP to RCP, respectively. Figure~\ref{fig:linear}(a) schematically illustrates these polarization-resolved components. The nonlinear circular dichroism $\mathrm{CD}^{(3 \omega)}$ is defined as
\begin{equation}\label{eq:CD_nonlin}
    \mathrm{CD}^{(3 \omega)} = \frac{I^{(3 \omega)}_{\text{RR}} - I^{(3 \omega)}_{\text{LL}}}{I^{(3 \omega)}_{\text{RR}} + I^{(3 \omega)}_{\text{LL}}},
\end{equation}
where $I^{(3 \omega)}$ is the third-harmonic intensity, and the first (second) subscript denotes the corresponding pump (generated harmonic) polarization state.

By varying $\alpha$ over its full range, we control the symmetry of the metasurface and, consequently, its linear and nonlinear CD. For the rotation angles at which the axis of the elliptical meta-atom coincides with a mirror symmetry axis of the lattice [see Figure~\ref{fig:metasurface_design}(c)], both linear and nonlinear CDs vanish due to the presence of in-plane mirror symmetry~\cite{koshelev_scattering_2024}. For intermediate orientation angles, where the ellipses are rotated away from the mirror-symmetric directions, the in-plane mirror symmetry is broken. While linear co-polarized circular dichroism is still forbidden due to the preserved z-symmetry plane of the membrane, total linear circular dichroism arises due to polarization conversion process. Strikingly, as we show further, the in-plane symmetry breaking induces strong nonlinear dichroism, which approaches unity for a specific orientation of the ellipses. Furthermore, for complementary angles, both total linear CD and nonlinear CD are expected to have opposite signs due to symmetry considerations.

\subsection{Linear optical response}

\begin{figure*}
\begin{center}
\includegraphics[width=0.99\linewidth]{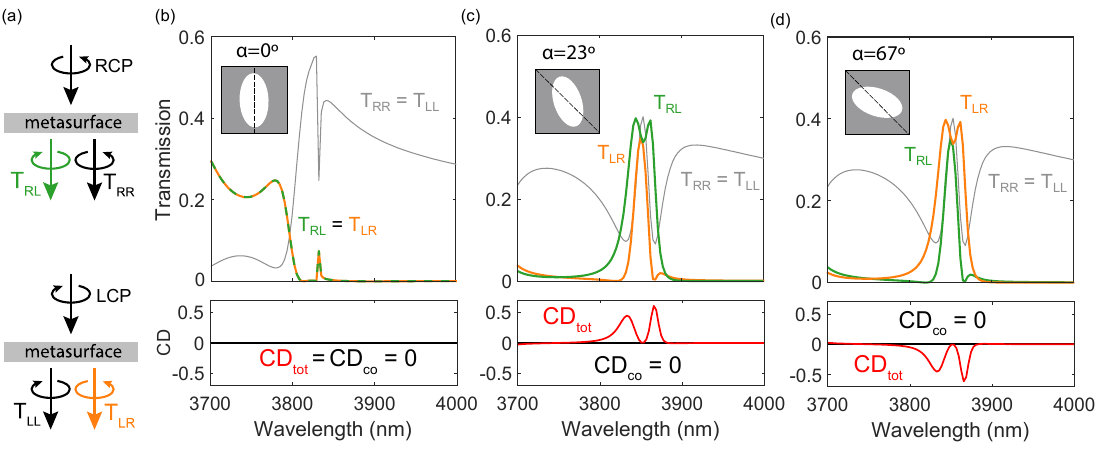}
\caption{\textbf{Linear circular dichroism through in-plane symmetry breaking.}
(a)~Schematic of circularly polarized transmission channels: right- (RCP) or left-circular (LCP) input decomposes into co-polarized ($T_\mathrm{RR}$, $T_\mathrm{LL}$) and cross-polarized ($T_\mathrm{RL}$, $T_\mathrm{LR}$) components.
(b)--(d)~Simulated transmission spectra (top) and circular dichroism (bottom) for ellipse rotation angles $\alpha = 0^\circ$, $23^\circ$, and $67^\circ$, respectively.
At $\alpha=0^\circ$, mirror symmetry enforces $T_\mathrm{RL}=T_\mathrm{LR}$ and $T_\mathrm{RR}=T_\mathrm{LL}$, yielding both ${CD}_{\text{co}}$ and ${CD}_{\text{tot}}$ to be zero.
Breaking in-plane symmetry allows conversion circular dichroism, while $\mathrm{CD}_\mathrm{co}$ remains zero.
The sign of $\mathrm{CD}_\mathrm{tot}$ reverses between $\alpha=23^\circ$ and $67^\circ$, consistent with symmetry arguments.}
\label{fig:linear}
\end{center}
\end{figure*}

To analyse the metasurface chirality, we numerically simulated its performance in the linear regime. Figure~\ref{fig:linear} shows the simulated transmission spectra for metasurfaces with meta-atom orientation angle $\alpha = \ang{0}$, $ \ang{23}$ and  $\ang{67}$. The presence of out-of-plane mirror symmetry, since the metasurface is a free-standing membrane, ensures that $T_{\text{RR}} = T_{\text{LL}}$~\cite{shalin_all-dielectric_2023}. The co-polarized transmission components for LCP ($T_{\text{LL}}$) and RCP ($T_{\text{RR}}$) are shown in the top panel of Figure~\ref{fig:linear}(b)--(d) in grey. For all three meta-atom orientations, these components coincide, consistent with the preserved out-of-plane mirror symmetry. In contrast, the cross-polarized transmission components, $T_{\text{LR}}$ and $T_{\text{RL}}$, displayed in orange and green, respectively, exhibit distinct behavior depending on $\alpha$.

To quantify the linear chiral response, we employ Eqs.~\eqref{eq:CD_lin}. The resulting co- and total circular dichroism, ${CD}_{\text{co}}$ and ${CD}_{\text{tot}}$, are showm in the bottom panel of Figures~\ref{fig:linear}(b)--(d). For the symmetric metasurface with $\alpha = \ang{0}$, both ${CD}_{\text{co}}$ and ${CD}_{\text{tot}}$ vanish across the entire wavelength range [bottom panel of Figure~\ref{fig:linear}(b)], as expected from symmetry considerations~\cite{koshelev_scattering_2024}. In contrast, for orientations that break the in-plane symmetry, such as $\alpha=\ang{23}$ and $\ang{67}$, the cross-polarized components differ, as shown in the top panel of Figure~\ref{fig:linear}(c) and Figure~\ref{fig:linear}(d), resulting in a non-zero ${CD}_{\text{tot}}$. Specifically, for $\alpha = \ang{23}$, ${CD}_{\text{tot}}$ deviates from zero in the vicinity of the metasurface resonances, reaching values of $0.45$ and $0.61$. For the complementary angle of $\alpha = \ang{67}$, ${CD}_{\text{tot}}$ reverses sign, attaining values of $-0.45$ and $-0.61$.

Eigenmode analysis further supports these observations. As detailed in section S2 of the Supplementary Material, two spectrally close modes are identified in simulations, consistent with the features observed in transmission spectra. The corresponding electric field profiles reveal that both modes are predominantly localized within silicon regions. This strong field confinement is particularly advantageous for enhancing nonlinear optical processes, where high local field intensities play a crucial role.

Next, we fabricated silicon free-standing membrane metasurfaces using the simulated parameters, with meta-atom orientation angles ranging from 0 to $\ang{90}$ in steps of $\ang{1.5}$. The fabricated metasurfaces exhibit behaviour consistent with the simulations. Due to the experimental limitations, we are only able to measure the overall LCP and RCP responses, without resolving the individual LCP and RCP components in the transmitted signal. The resulting transmission spectra are shown in Section S1 of the Supplementary Material.
In the experimental spectra,  only a single pronounced resonance is observed, arising from two closely spaced modes which cannot be distinguished due to the absence of decomposition of the transmitted signal into its LCP and RCP components. The CD for  $\alpha = \ang{0}$ is close to zero, with slight deviations attributed to the non-ideal fabrication and limitation of the optical setup. For complementary angles  $\alpha = \ang{23}$ and  $\alpha = \ang{67}$, the CD sign reverses across the resonance. 

\begin{figure*}
    \begin{center}
    \includegraphics[width=0.7\linewidth]{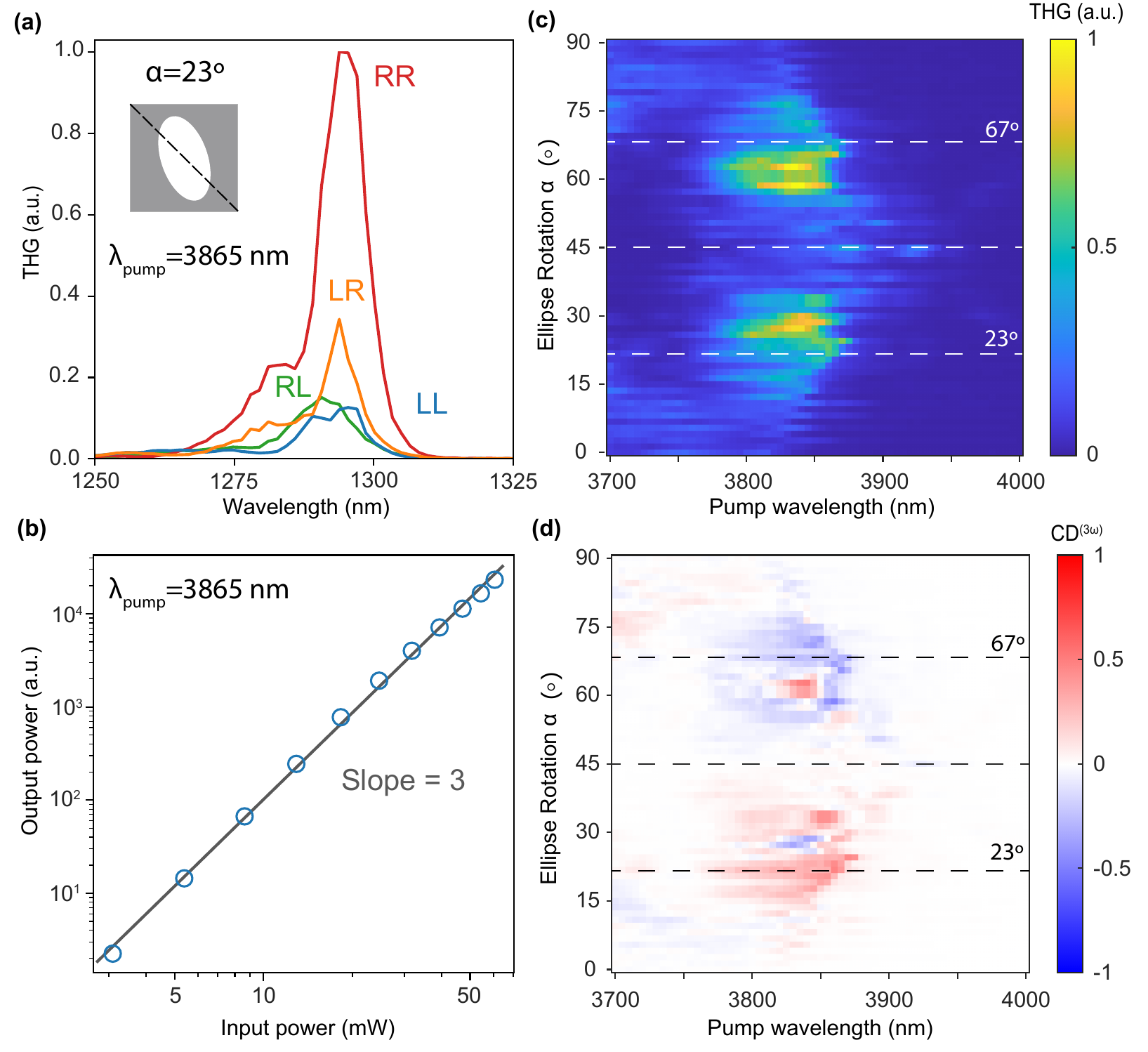}
    \caption{ \textbf{Chiral third harmonic generation (experiment).}  (a) Third-harmonic generation spectra from a membrane with $\alpha=23^\circ$ and a pump wavelength 3865~nm, resolved into its input and output circular polarization. (b) Power dependence of the peak third-harmonic generation at the point of signal maximum. (c) Peak RCP-RCP third-harmonic signal and (d) $\text{CD}^{(3 \omega)}$ as functions of pump wavelength and meta-atom rotation angle $\alpha$. The CD map opacity is scaled by the third-harmonic signal at each point to suppress noise in the regions of low signal.}
    \label{fig:nonlinear1}
    \end{center}
\end{figure*}

We further examine the linear optical response of the metasurfaces as a function of meta-atom orientation angle over the full range from $\ang{0}$ to $\ang{90}$. The simulated and experimental transmission spectra for RCP and LCP exhibit consistent behaviour (see Supplementary Figure S2). As angle $\alpha$ increases, the resonant mode undergoes a progressive redshift up to $\alpha = \ang{45}$. Beyond this point, the resonance exhibits a blueshift, returning to its original spectral position at $\alpha = \ang{90}$. This behaviour arises from the symmetry of the structure, as the metasurface at $90^\circ$ is equivalent to that at $0^\circ$. A similar trend is observed in total linear CD (see Supplementary Figure S3). Numerical simulations reveal a clear sign reversal of the CD upon crossing the meta-atoms orientation angle of $45^\circ$. The experimental results follow the same overall tendency; however, the resonant features are less pronounced, likely due to fabrication imperfections and measurement limitations.

\subsection{Nonlinear optical response}

We now turn to the nonlinear optical properties of the membrane, focusing on the third-harmonic generation (THG) process. Each metasurface was pumped with LCP and RCP femtosecond excitation in the wavelength range of 3700-4000~nm, and the transmitted THG signal was resolved into its circular polarization components (see Methods and Supplementary Section S3 for more detail). Figure~\ref{fig:nonlinear1}(a) shows the polarization-resolved transmitted THG spectra at the wavelength corresponding to the maximum output harmonic intensity, obtained from the membrane with $\alpha = 23^\circ$ under excitation at 3865~nm. The dominant contribution arises from the $I^{(3 \omega)}_{\text{RR}}$ component, while the remaining channels exhibit comparatively weaker responses. The observed double-peak structure likely originates from the finite linewidth of the excitation laser and imperfect mode overlap, and can also be attributed to the presence of a second resonant mode as observed in the simulations (Section S2 of the Supplementary Material). Figure~\ref{fig:nonlinear1}(b) presents the dependence of the THG on the input power, exhibiting a clear cubic scaling that confirms the third-order nonlinear origin of the process.

    \begin{figure*}
    \begin{center}

  \includegraphics[width=0.99\linewidth]{Figures/fig4.pdf}
  
  \caption{ \textbf{Maximum nonlinear chirality from the membrane metasurfaces (experiment).} Third-harmonic generation resolved into circular polarization components for angles (a) $\alpha = 0^\circ$, (b) $\alpha = 23^\circ$, and (c) $\alpha = 67^\circ$. The inset shows the orientation of the meta-atoms. The third-harmonic generation signal is normalised to the maximum of each plot. Circular dichroism in the high-THG region near structural resonance is shown for (d) $\alpha = 0^\circ$, (e) $\alpha = 23^\circ$ and (f) $\alpha = 67^\circ$. The inset shows SEM of the metasurfaces with the corresponding meta-atoms orientations. The scale bar is 1 µm.}
  \label{fig:nonlinear_CD}
  \end{center}

\end{figure*}

Next, we examine the THG response of metasurfaces with varying meta-atom orientation angles. Figure~\ref{fig:nonlinear1}(c) shows the peak THG intensity for metasurfaces with $\alpha$ ranging from $\ang{0}$ to $\ang{90}$ in steps of $\ang{1.5}$, under excitation at pump wavelengths between 3700 and 4000~nm. The maximum THG intensity occurs at $23^\circ/67^\circ$ and a pump wavelength of 3850~nm. The THG response is symmetric with respect to $\alpha = \ang{45}$, and the spectral position of the maximal enhancement follows the linear resonances (see Section S1 in the Supplementary Material), confirming its resonant origin. The stronger THG enhancement at specific angles, rather than a uniform increase following the resonance positions observed in transmission spectra, is attributed to the variation of the $Q$-factor with $\alpha$ and the resulting differences in nonlinear coupling efficiency to the resonant modes\cite{tonkaev_unconventional_2025}. The THG signal was further resolved into LCP and RCP components, and the nonlinear CD was calculated using Eq.~\ref{eq:CD_nonlin}. Figure~\ref{fig:nonlinear1}(d) presents the resulting $\text{CD}^{(3 \omega)}$ as a function of $\alpha$ and pump wavelength. For all pump wavelengths, the nonlinear CD changes sign across $\alpha = 45^\circ$, in agreement with symmetry considerations. The maximum chiral response coincides with the peak THG intensity at $23^\circ/67^\circ$ and 3850~nm, highlighting the role of resonant modes in enhancing both the chiral and nonlinear response. Interestingly, near $\alpha = 30^\circ/60^\circ$ at 3850~nm, the nonlinear CD reverses sign over a narrow spectral and angular range. This behavior can be attributed to the resonant metasurface modes at the THG frequency, as no particular features are observed in this parameter range neither in linear transmission spectra nor in mode dispersion diagrams at the pump wavelengths (see Supplementary Figures S2-S4).

Finally, to gain deeper insight into the origin of the enhanced nonlinear chirality, we examine in detail the regions of high nonlinear CD identified above. We begin with the metasurface possessing in-plane symmetry ($\alpha = \ang{0}$), for which the nonlinear CD is expected to vanish. Figure~\ref{fig:nonlinear_CD}(a) shows the peak THG intensity as a function of the pump wavelength, resolved into circular polarization components. The co-polarized RCP and LCP components dominate and are shown in red and blue, respectively, while the cross-polarized components corresponding to the conversion of RCP to LCP and LCP to RCP are shown in green and orange, respectively. The co- and cross-polarized contributions are nearly identical ($I_\mathrm{LL} = I_\mathrm{RR}$ and $I_\mathrm{LR} = I_\mathrm{RL}$, respectively), indicating the absence of chiral selectivity. Nonlinear CD calculated from these data using Eq.~\ref{eq:CD_nonlin} is shown in Figure~\ref{fig:nonlinear_CD}(d) and confirms that the nonlinear CD remains close to zero irrespective of the resonant behaviour of the system, consistent with the symmetry arguments.

We then consider the metasurfaces with $\alpha = \ang{23}$ and $\alpha = \ang{67}$, which exhibit the strongest nonlinear CD. Figures~\ref{fig:nonlinear_CD}(b,c) show the THG response resolved into circular polarization components. In both cases, a clear asymmetry between the components emerges, with the co-polarized channel dominating:  RR for $\alpha = 23^\circ$ and LL for $\alpha = 67^\circ$. These dominant components exhibit a pronounced resonant enhancement centred at 3840~nm. Notably, the responses for the complementary angles show strong correspondence in both absolute and relative intensities, reflecting the underlying symmetry of the structure. Figures~\ref{fig:nonlinear_CD}(e,f) present the corresponding nonlinear CD in the high-THG region for $\alpha = 23^\circ$ and $\alpha = 67^\circ$, respectively. The nonlinear CD reaches peak values of $0.79$  and $-0.81$ for the two complementary angles. While the sign of the CD is predominantly governed by the structural symmetry, the resonant modes play a crucial role in increasing the THG efficiency and enhancing the visibility of the nonlinear CD response. 

For comparison, we also evaluate THG from the unpatterned silicon free-standing membrane. In this case, the signal is almost three orders of magnitude weaker than that of the metasurface at the resonance and exhibits no chiral response. Instead, the emission is dominated by cross-polarized components, while co-polarized contributions are nearly suppressed (see Section S4 in the Supplementary Material). This behaviour arises from the intrinsic symmetry of the silicon atomic lattice in the absence of the resonant structuring. In particular, the C$_4$ symmetry of the lattice permits only cross-polarized THG, whereas co-polarized contributions are symmetry-forbidden~\cite{koshelev_scattering_2024}.

\section{Conclusions}
We have investigated systematically both linear and nonlinear chiral properties of free-standing dielectric membrane metasurfaces composed of meta-atoms with variable orientation angles. We have demonstrated strong co-polarized circular dichroism without breaking the out-of-plane mirror-reflection symmetry. In particular, the metasurface supported a chiral nonlinear resonance that simultaneously enhanced both the linear and nonlinear responses. Under these conditions, the structure exhibited strong THG enhanced more than two orders of magnitude compared with an unpatterned membrane, while the circular dichroism could be tuned from $-0.81$ to $+0.79$ through simple rotation of the meta-atoms. These findings establish a novel approach to engineering nonlinear chiral optical responses in planar metasurfaces, and open opportunities for compact photonic platforms for selective chiral harmonic generation, nonlinear polarization control, and application in ultrathin chiral light sources, optical information processing, and enantiomer-selective spectroscopy.

\section{Methods}
\subsection*{Fabrication} The membrane was fabricated in a silicon-on-insulator (SOI) wafer with (100) orientation comprising a 1~$\mu$m boron-doped silicon device layer ($8.5$--$11.5$~$\Omega$\,cm), a 1~$\mu$m silica buried oxide layer, and a 250~\textmu m Si handle substrate. To define the membrane regions, a 3~\textmu m silica layer was deposited on the wafer backside by PECVD (Plasma-Therm Corial D250L), patterned by direct laser writing (Heidelberg MLA 150, AZ ECI 3027 at 3000~rpm), and etched using fluorine based DRIE (SPTS Advanced Plasma System).The membrane patterns were written using electron-beam lithography (Raith EBPG5000, ZEP520 resist diluted $50\%$, spun 2000~rpm), and transferred by DRIE (Alcatel AMS 200 SE) into the silicon. The marked backside openings were subsequently etched through the handle layer via Bosch DRIE, exposing the buried oxide, which was selectively removed by HF vapour etching (SPTS Etch) to release the suspended membranes.In total, 64 membranes were fabricated, each with elliptical holes rotated from $0^\circ$ to $90^\circ$ in $1.5^\circ$ steps.

\subsection*{Linear measurements}
The mid-infrared transmission spectra at normal incidence were obtained using a Bruker Vertex 80v FT-spectrometer with an attached IR Microscope (HYPERION 3000) equipped with a
liquid nitrogen-cooled MCT detector. The metasurfaces were excited using a ZnSe lens with a focal length of 50 mm, mildly focusing IR light on the sample surface. Polarization was
controlled by a motorised quarter-waveplate (QWP, B.Halle, 2.5-7~\textmu m, MgF2). Transmitted light was collected with a 25~mm lens equipped with an additional iris placed at its back focal plane. Closing the iris allowed for limiting the numerical aperture of the system down to approximately 0.02 and thus suppressing the unwanted signal from oblique excitation angles. The signal collection area was limited to an approximately 300~\textmu m square central region of the membrane by a double-blade aperture placed in the conjugate image plane of the IR microscope. The sample chamber was constantly purged with dry air to provide a stable low level of humidity.

\subsection*{Nonlinear measurements} An experimental diagram is presented in the Supporting Information (Figure S7). The light source used was an Ekspla Femtolux 3 femtosecond laser, operating at a wavelength of 1030~nm with a pulse duration of 250~fs and a repetition rate of 1.15~MHz. The output wavelength of the laser was changed to the mid-IR regime (3700-4000~nm) using an optical parametric amplifier (MIROPA from Hotlight Systems). This was transformed to circular polarization using an achromatic quarter-waveplate (B.Halle, 3-6~\textmu m), before being focused onto the metasurface using a lens (Thorlabs LA5763). The nonlinear output from the metasurface was collimated using an objective lens (Mitutoyo Plan Apo NIR 20x) to ensure maximum first-order collection. This output was passed through a quarter-waveplate (Thorlabs AQWP05M-1600) and linear polarizer (Thorlabs WP25M-UB1), allowing us to extract specific output circular polarizations. The intensity of each third-harmonic component was measured using an NIR spectrometer (Ocean Optics NIRQuest).

\medskip
\textbf{Acknowledgements} \par 
Y.K. acknowledges a support from the Australian Research Council (Grant No. DP210101292) and the International Technology Center Indo-Pacific (ITC IPAC) via Army Research Office (contract FA520923C0023). F.U.B., I.S., and H.A. thank the Swiss State Secretariat for Education, Research and Innovation (SERI) for financial support under the contract numbers 22.00018 and 22.00081 in connection with the projects from the European Union’s Horizon Europe Research and Innovation Program under agreements 101046424 (TwistedNano) and 101070700 (MIRAQLS).

\medskip

\bibliographystyle{naturemag}
\bibliography{Refs}

\end{document}